\documentclass[pra,twocolumn,floatfix,aps]{revtex4}
\usepackage{amsmath}
\usepackage{makeidx}
\usepackage{amssymb}
\usepackage{graphicx}

\usepackage[usenames]{color}
\usepackage[normalem]{ulem}

\usepackage{hyperref}
\usepackage[T1]{fontenc} 

\begin{document}

\title{Multi-ring, stripe, and super-lattice   solitons in a spin-orbit coupled 
spin-1 condensate}

\author{S. K. Adhikari\footnote{sk.adhikari@unesp.br      \\  https://professores.ift.unesp.br/sk.adhikari/ }}

\affiliation{Instituto de F\'{\i}sica Te\'orica, Universidade Estadual Paulista - UNESP, 01.140-070 S\~ao Paulo, S\~ao Paulo, Brazil}
      
%%%%%%%%%%%%%%%%%%%%%%%%%%%%%%%%%%%%%%%%%%%%%%%%%%%%%%%%%%%%%%%%%%%%%%%%%%%%%%
%%%%%%%%%%                    Abstract                             %%%%%%%%%%%
%%%%%%%%%%%%%%%%%%%%%%%%%%%%%%%%%%%%%%%%%%%%%%%%%%%%%%%%%%%%%%%%%%%%%%%%%%%%%%

\date{\today}

\begin{abstract}

We demonstrate  exotic stable quasi-two-dimensional solitons  
in a Rashba or a Dresselhaus  spin-orbit (SO) coupled  hyperfine 
spin-1 ($F=1$) trap-less antiferromagnetic
Bose-Einstein condensate    using  the mean-field Gross-Pitaevskii equation. 
For weak SO coupling, the solitons are of the $(-1,0,+1)$ or $(+1,0,-1)$ type with intrinsic vorticity,
for Rashba or Dresselhaus SO coupling, 
where the numbers in the parentheses  denote  angular momentum in spin components $F_z= +1,0,-1$, respectively. 
For intermediate SO coupling, the solitons have multi-ring structure maintaining the above-mentioned vortices  in respective components. For larger SO coupling, {\it super-lattice} solitons with  a square-lattice structure in total density 
are found in addition to stripe solitons with stripe pattern in component densities with no periodic modulation in total density.

\end{abstract}

 \maketitle

A bright soliton or a soliton is  a one-dimensional (1D) shape-preserving solitary wave  stabilized by a cancellation of dispersive forces and nonlinear attraction  \cite{r1}.   
  Quasi-1D solitons have been observed in a  Bose-Einstein condensate (BEC)
of $^7$Li \cite{r2} and $^{85}$Rb \cite{r3} atoms following a theoretical suggestion \cite{r5}.
%based on  the mean-field Gross-Pitaevskii (GP) equation \cite{r6}. 
%Solitons have also been studied in binary BEC mixtures \cite{r7}. 
  However, solitons cannot be stabilized in two (2D) \cite{townes} and three dimensions (3D) \cite{r1}. 

The interest in the creation of  spin-orbit (SO) coupling in a   BEC   increased  after the 
observation of
a hyperfine
spin-1 spinor BEC of $^{23}$Na atoms \cite{exptspinor}.   
%Spinor BECs can show a rich variety of topological excitation \cite{thspinor,kita} not possible in a scalar BEC. 
Although there cannot be a natural SO coupling in a neutral atom, it is  possible to introduce an artificial synthetic  SO coupling   by tuned   
Raman lasers coupling  the different spin states 
\cite{thso,exptsob}. 
Two such possible SO couplings are due to Dresselhaus \cite{SOdre} and  Rashba \cite{SOras}. 
An equal mixture of 
Dresselhaus  and Rashba SO couplings has been experimentally realized 
in a pseudo spin-1/2 $^{87}$Rb  \cite{exptso}
and  $^{23}$Na \cite{na-solid}
BEC of  $F_z=0,-1$
spin states. 
%Possible ways of realizing the SO coupling in a three-component   spin-1 BEC have
%been considered \cite{3c}. 
% which is a simplification over the three-component spin-1 hyperfine spin state 5S$_{1/2}$ of $^{87}$Rb.   
 Later, % pseudo spin-1/2 SO-coupled BECs were   studied  experimentally \cite{olab} and theoretically \cite{exptsob} and 
an SO-coupled 
 spin-1 $^{87}$Rb BEC of $F_z =\pm1,0$ states was  observed  \cite{exptsp1}.
%A three-component SO-coupled spin-1 BEC is  known to exhibit a rich variety of physical phenomena \cite{thspinor}
%not possible in a two-component pseudo spin-1/2 BEC. 

A pseudo spin-1/2 SO-coupled  BEC can sustain a quasi-1D \cite{quasi-1d} or a quasi-2D \cite{quasi-2d} or a   3D  \cite{quasi-3d} soliton.   In a spin-1/2 SO-coupled BEC, quantum half-vortex solitons   \cite{half-vortex}  have also been studied  and   multi-ring and rotating solitons were also found   in a  periodic potential \cite{radper}. 
Stability of these solitons have also been studied recently \cite{stab}.
% All these studies were confined to  two-component pseudo spin-1/2 SO-coupled BEC. 
 A   {spin-1} SO-coupled BEC also 
sustains a quasi-1D \cite{quasi-1d1} or a quasi-2D \cite{quasi-2d1} or a 3D 
\cite{quasi-3d1} soliton.  
% Stripe and other structures in density have been found in a Rashba coupled pseudo spin-1/2 and spin-1 
%BEC \cite{referee1}.

Here we undertake  a comprehensive study of solitons in a spin-1 
 quasi-2D { Rashba or a Dresselhaus } SO-coupled {antiferromagnetic} BEC.
For a weakly  Rashba or Dresselhaus
 SO-coupled  spin-1    BEC (SO-coupling strength $\gamma \lessapprox 0.75$), the  soliton is of the 
 $(\mp 1,0,\pm 1)$ type 
of angular momenta
 $\mp 1,0,\pm 1$ in   components $F_z=+1,0,-1$ \cite{kita,quasi-2d1}, where the positive (negative) sign denotes a vortex (an antivortex) and the upper (lower) sign corresponds to Rashba (Dresselhaus) SO coupling. For medium SO-coupling,   we find  $(\mp 1,0,\pm 1)$-type
multi-ring solitons   for    Rashba or Dresselhaus SO couplings,  
 maintaining the 
above-mentioned vortices at the center of respective components. 
  For larger SO-coupling, two types of quasi-degenerate solitons are found. The first type 
has maxima and minima in density forming a stripe pattern. In the second type, matter is spontaneously distributed on a 2D square lattice.   {
For a Dresselhaus SO-coupled BEC, the density  and energy of the  
solitons are the same as in the 
case of Rashba coupling, but the two wave functions are different.      }
% The   formation of a super-lattice phase  without any external trap seems to be a 2D generalization of the theoretically predicted \cite{sprsld,stripe} and experimentally observed \cite{st2}    super-solid    stripe phase in an   SO-coupled spinor BEC. The spontaneous formation of a super-solid 
%{2D super-lattice} state \cite{sprsld} in {untrapped} atomic and condensed-matter systems was never 
%confirmed before despite a  few suggestions in solid helium \cite{solhel} and trapped atomic gases \cite{dipolar} with long-range interaction.

{The pursuit of a super-solid \cite{sprsld}
 is a fascinating subject in different areas of condensed matter physics. 
A super-solid  is a quantum state where matter forms a spatially-ordered rigid structure, { breaking continuous translational symmetry},  as in
a crystalline solid, and also enjoys friction-less flow  as in a super-fluid, { breaking Gauge symmetry.} Although, super-solidity was never confirmed  in super-fluid 
helium \cite{solhel}, more recently, 
it has been suggested \cite{losh}  and observed \cite{dipolar}  in a dipolar BEC. In a spin-1/2 SO-coupled BEC, super-solidity has been realized experimentally \cite{exp-stripe} in the form of a quasi-1D stripe pattern in density \cite{stripe,stripe2,sinha}.{ The present  
solitons with a   2D square-lattice structure in total density \cite{2020}, viz. Figs. \ref{fig5}(c) and (f) and Fig. \ref{fig4}(f),  sharing 
properties with a conventional supersolid,  
will be termed  super-lattice solitons  in the following as suggested in Refs. \cite{stripe,2020}.   The  super-lattice
solitons are expected to possess coexisting order parameters with super-solid-like properties leading to  a 
periodic spatial structure in  total density \cite{stripe,2020}, as illustrated in Fig. 1 of  Ref. \cite{2020}, 
 in Fig. 1 of Ref. \cite{stripe,sting} and in Fig. 1(e) of Ref. \cite{sinha,sinha2}.  The multi-ring, viz. Fig. \ref{fig3}, and stripe, viz. Figs. \ref{fig4}(a)-(c),   solitons only exhibit a periodic structure in the component densities, 
due to a phase separation among the  components 
without any periodic structure in the total density. }      
A super-lattice soliton 
is a unique  example of spontaneous 2D crystallization
in a free condensed matter system and  a study of this  may  enhance the knowledge about  the origin of  crystal formation in solids under controlled conditions. 
%Also, the evolution of the  1D super-stripe \cite{exp-stripe,exptsob}
%in a SO-coupled spin-1/2 BEC to a spin-1 SO-coupled  2D super-lattice is highly nontrivial that justifies and motivates this study.}

%For the study of soliton formation in  a  quasi-2D  spin-1 spinor BEC, 
We consider a BEC of $N$ atoms, each of mass { $m$}, under a
harmonic trap $V({\bf r})=  { m}\omega_z^2 z^2/2$  of  frequency $\omega_z$ in the $z$ direction and free in the $x-y$ plane. After integrating out the $z$ 
coordinate \cite{quasi12d}, the single particle Hamiltonian of the SO-coupled  BEC is
%
%The single-particle Hamiltonian 
%of the condensate without atomic interaction   in this quasi-2D trap in dimensionless  variables
 \cite{exptso} 
\begin{align}\label{sp}
H_0 =-\frac { \hbar^2}{2 m}  \nabla_{\boldsymbol \rho}^2 + \gamma [\eta p_y \Sigma_x -  p_x \Sigma_y],
\end{align}
%\begin{equation}
%H_0 = \frac{p_x^2+p_y^2}{2{  m}}  +H_{\mathrm{SO}}  ,
%\label{sph} 
%\end{equation}
where ${\boldsymbol \rho}\equiv \{x,y   \}$, {$\nabla_{\boldsymbol \rho}^2=
(\partial^2/\partial x^2+\partial^2/\partial y^2) \equiv (\partial_x^2+\partial_y^2)$},
$\gamma$ is the strength of the SO-coupling term in square bracket,
%where %${\bf r}=\{x,y,z\}$, $\nabla^2_{\bf r}=p_x^2+p_y^2+p_z^2,$
%$p_x = -i\hbar \partial_x \equiv -i\hbar \partial/\partial x, p_y = -i\hbar \partial_y \equiv -i\hbar %\partial/\partial y $. 
%are the momentum operators along $x$ and $y$ axes, respectively,  .
{where $\eta=+1$
for Rashba coupling,  $\eta =-1$ for Dresselhaus coupling, { and $\eta=0$ for an equal mixture of 
Rashba and Dresselhaus couplings,}
%{multi For Rashba and Dresselhaus SO couplings, respectively, the SO-coupling term } \cite{exptso}
%\begin{align}
%$H_{\mathrm{SO}}= - \gamma p_x \Sigma_y+\gamma p_y \Sigma_x,$ and
%$$H_{\mathrm{SO}}=  - \gamma p_x \Sigma_y-\gamma p_y \Sigma_x,$
%\end{align}
 $p_x=-i\hbar \partial_x, p_y=-i\hbar \partial_y$ and }
the spin matrices  $\Sigma_x$ and $\Sigma_y$ are
\begin{eqnarray}
\Sigma_x=\frac{1}{\sqrt 2} \begin{pmatrix}
0 & 1 & 0 \\
1 & 0  & 1\\
0 & 1 & 0
\end{pmatrix}, \quad  \Sigma_y=\frac{i}{\sqrt 2 } \begin{pmatrix}
0 & -1 & 0 \\
1 & 0  & -1\\
0 & 1 & 0
\end{pmatrix}.
\end{eqnarray}
%and $\gamma$ is the strength of SO coupling.  
%The numerical results for density and energy of the solitons are the same for both types of SO coupling, although the respective wave functions are different.

A quasi-2D Rashba SO-coupled  spin-1 spinor  BEC is described by the 
 following set of  Gross-Pitaevskii equations \cite{r6} at zero temperature  for spin components $F_z = \pm 1, 0$ \cite{thspinor,thspinorb}
\begin{align}\label{EQ1} 
i \partial_t &\psi_{\pm 1}({\boldsymbol 
\rho})= \left[{\cal H}+{c_2}
\left(n_{\pm 1} -n_{\mp 1} +n_0\right)  \right] \psi_{\pm 1}({\boldsymbol 
\rho})\nonumber \\
+&\left\{c_2 \psi_0^2({\boldsymbol \rho})\psi_{\mp 1}^*({\boldsymbol \rho})\right\} % \nonumber \\
-i {\widetilde \gamma} (\eta\partial_y \pm i \partial_x)  \psi_{0}  ({\boldsymbol \rho})  \, , 
\\ \label{EQ2}
i \partial_t&\psi_0({\boldsymbol \rho})=\left[ {\cal H}+{c_2}
\left(n_{+ 1}+n_{- 1}\right) \right] \psi_{0}({\boldsymbol \rho})+\big \{2c_2 \psi_{+1}({\boldsymbol \rho}) \nonumber \\
\times &\psi_{-1}({\boldsymbol \rho})\psi_{0}^* ({\boldsymbol \rho})\big\}   
{-i{\widetilde \gamma} [-i  \partial_x \phi_{-1}  ({\boldsymbol \rho})  + \eta \partial_y \phi_{+1} ({\boldsymbol \rho}) ]}  
  \, , \\
{\cal H}=&-\frac{1}{2}\nabla^2_{\boldsymbol \rho}+ c_0 n,    \\
c_0 =&\frac{2{N}\sqrt{2\pi}(a_0+2a_2)}{3}, \quad c_2 
= \frac{2{N}\sqrt{2\pi}(a_2-a_0)}{3}, \label{EQ4}
\end{align}
where { $\phi_{\pm 1}({\boldsymbol \rho})=\psi_{+1}({\boldsymbol \rho})\pm \psi_{-1}({\boldsymbol \rho})$,}
 %${\boldsymbol \rho}\equiv \{x,y   \}$, $\nabla_{\boldsymbol \rho}^2\equiv (\partial_x^2+\partial_y^2)$, 
   $\partial_t \equiv \partial/\partial t$,  
 $\widetilde \gamma = \gamma/\sqrt 2 $, $n_j = |\psi_j|^2, j=\pm 1,0$ are the densities of spin components $F_z= \pm 1 , 0$, and $n ({\boldsymbol \rho})= \sum_j n_j({\boldsymbol \rho})$  the total density,    $a_0$ and
$a_2$ are the scattering lengths in the total spin 0
and 2 channels, respectively, and the asterisk denotes complex conjugate. All quantities in Eqs. (\ref{EQ1})-(\ref{EQ4}) and in the following are dimensionless; this is achieved by expressing length ($a_0,a_2,x,y,z$) in units of   oscillator length  
$l_0\equiv \sqrt{\hbar/{ m}\omega_z}$,
density  in units of $l_0^{-2}$, energy in units of $\hbar \omega_z$, and time in units of $\omega_z^{-1}$. 
 A spin-1 spinor BEC is   classified into two magnetic phases \cite{thspinor}:
ferromagnetic ($c_2<0$) and antiferromagnetic ($c_2>0$).
%The time dependence of the functions is  suppressed in Eqs.~(\ref{EQ1}) and (\ref{EQ2}).
 The normalization condition is 
%\begin{align}\label{noma}
$ {\textstyle \int} n({\boldsymbol \rho})\, d{\boldsymbol \rho}=1\, .$
%\end{align}
The time-independent version of Eqs.
 (\ref{EQ1})-(\ref{EQ2}), appropriate for stationary solutions, can be derived from the energy functional  
\begin{align}\label{energy}
E[\psi] &=  \frac{1}{2} \int d{\boldsymbol \rho} \Big\{ \sum_j |\nabla_{\boldsymbol \rho}\psi_j|^ 2
{-2\mu n}
+c_0n^2  + c_2\big[n_{+1}^2 
\nonumber \\  &   %+ c_2\big[n_{+1}^2 +n_{-1}^2
+n_{-1}^2+2(n_{+1}n_0+n_{-1}n_0-n_{+1}n_{-1}+\psi_{-1}^*\psi_0^2\psi_{+1}^* 
\nonumber \\ &  
+ \psi_{-1}\psi_0^{*2}\psi_{+1})  \big] -2i\widetilde \gamma\big[\eta \psi_0^* \partial_y\phi_{+1}
+\eta \phi_{+1}^*  \partial_y  \psi_0
\nonumber \\
&   
 -i \psi_0^* \partial_x\phi_{-1}+i \phi_{-1}^* \partial_x \psi_0 \big]  \Big\}{ + \mu},
\end{align}
where $\mu$ is the chemical potential.
%For Dresselhaus SO coupling  $\partial_y$ in     Eqs. (\ref{EQ1}), (\ref{EQ2}), and (\ref{energy}) should be replaced by $-\partial_y$.

\begin{figure}[!t] 
\centering
\includegraphics[width=.8\linewidth]{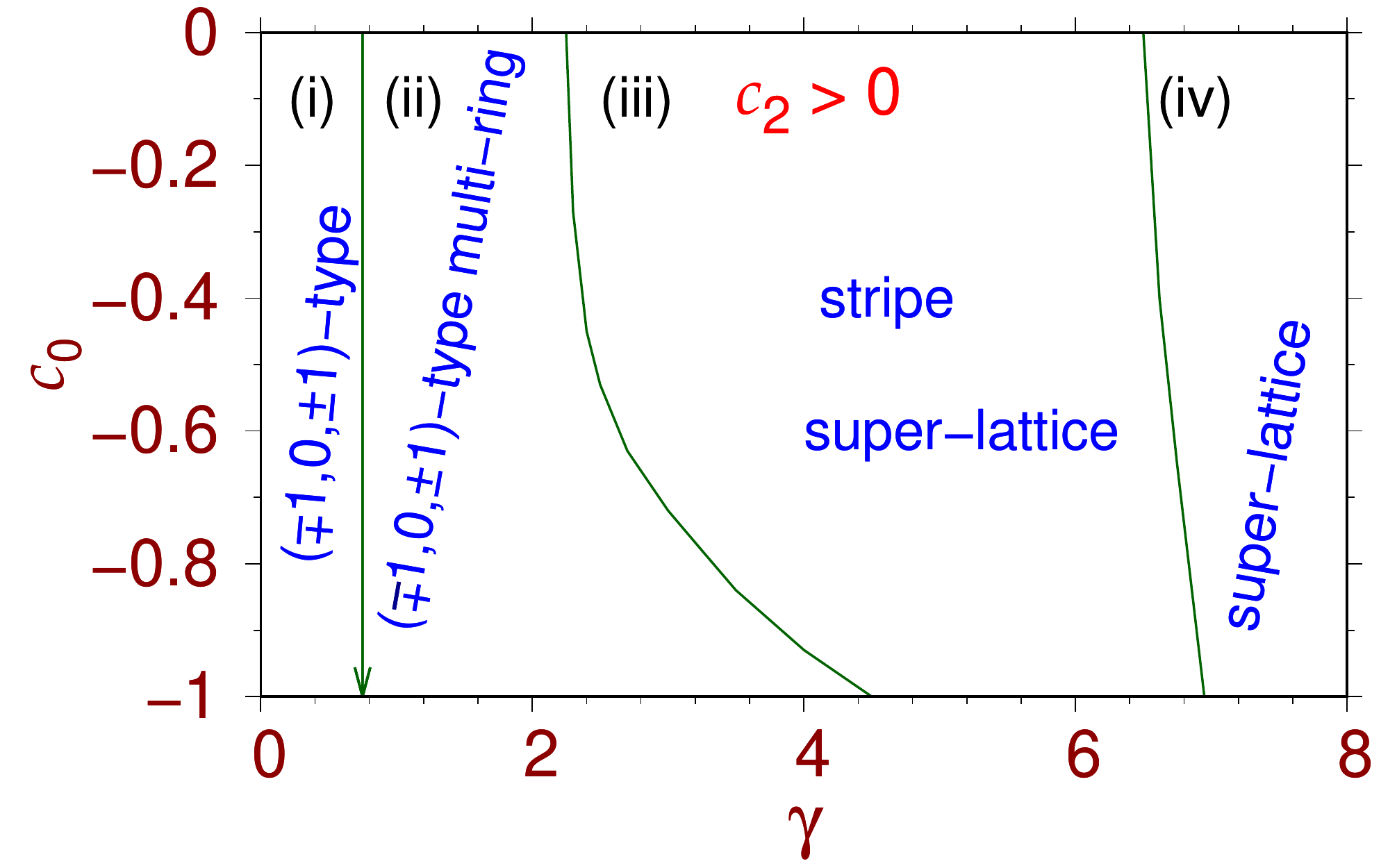}  
 
\caption{ The $c_0$ versus $\gamma$  phase plot showing { soliton formation in a Rashba or a Dresselhaus
SO-coupled   spin-1 {antiferromagnetic} BEC} in different regions of parameter space: formation of  (i) ($\mp 1,0,\pm 1$)-type,  (ii)  ($\mp 1,0,\pm 1$)-type
multi-ring,  (iii) stripe,  and (iii,iv)  super-lattice solitons. 
%For a fixed $c_0$ and $\gamma$, soliton formation is possible for $c_2>c_2^{(\mathrm{cr})}$. 
The upper (lower) sign correspond to Rashba (Dresselhaus) coupling. 
Results in all figures are plotted in dimensionless units. }
\label{fig1}

\end{figure}

To solve Eqs. (\ref{EQ1}) and (\ref{EQ2}) numerically, we propagate
these  in time by the split-time-step Crank-Nicolson discretization scheme \cite{bec2009} with {the boundary condition that the wave-funcion components and their first derivatives vanish at the boundary} while 
 using a space
step of 0.1 and a time step of 0.001 for imaginary-time propagation and 0.00025 for real-time 
propagation. { All calculations of soliton profiles employ  imaginary-time approach {with the 
conservation of normalization during time propagation,}
which 
finds the lowest-energy solution of each type. Real-time propagation is used to test the stability of the solitons. { Magnetization ($=\int d {\boldsymbol \rho[n_{+1}-n_{-1}] }$) is
not a good quantum number and is left to freely evolve
during time propagation to attain a final converged value
consistent with the parameters of the problem. However, this converged value was  found to be zero in all cases studied.} 
}
% respectively, to obtain the stationary state and to study the
%dynamics. % There are different C and FORTRAN programs for solving the GP equation \cite{bec2009,bec2012}  and one
%should use the appropriate one.   

We study the formation of  a quasi-2D soliton in a self-attractive $(c_0<0)$
spin-1 {antiferromagnetic}  BEC  for different sets of parameters: the nonlinearities $c_0,  c_2$ and SO-coupling strength $\gamma$. 
The scenario of soliton formation {for Rashba  or Dresselhaus SO coupling} is illustrated in the phase plot of  $c_0$ versus $\gamma$ for   $c_2>0$ in Fig. \ref{fig1}. Solitons of type $(\mp 1,0,\pm 1)$  are formed in region (i), whereas $(\mp 1,0,\pm 1)$-type
multi-ring solitons   are formed in region (ii). In region (iii), stripe   solitons are formed 
and  super-lattice solitons  are possible in regions (iii) and  (iv).
The phase plot remains unchanged  independent of the value of $c_2$ unless it is too large. {No visible change in the phase plot is found for $10>c_2>0$, consistent with an analytical result that the density 
and energy of a weakly SO-coupled spin-1 BEC soliton are independent of $c_2$ for  quasi-1D \cite{quasi-1d12} and  quasi-2D  \cite{quasi-2d1} settings.}
 For smaller values of $c_0$ ($c_0 < -1$), because of  an excess of attraction the soliton acquires a small size of the order of the lattice period and the crystalline structure is lost.

\begin{figure}[!t] 
\centering
\includegraphics[width=.325\linewidth]{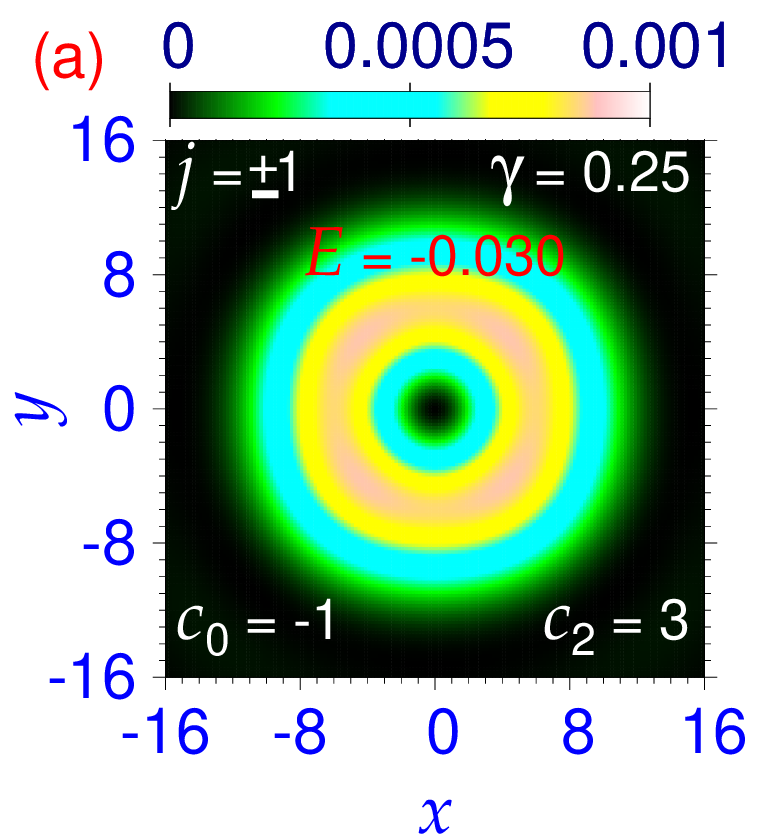} 
\includegraphics[width=.325\linewidth]{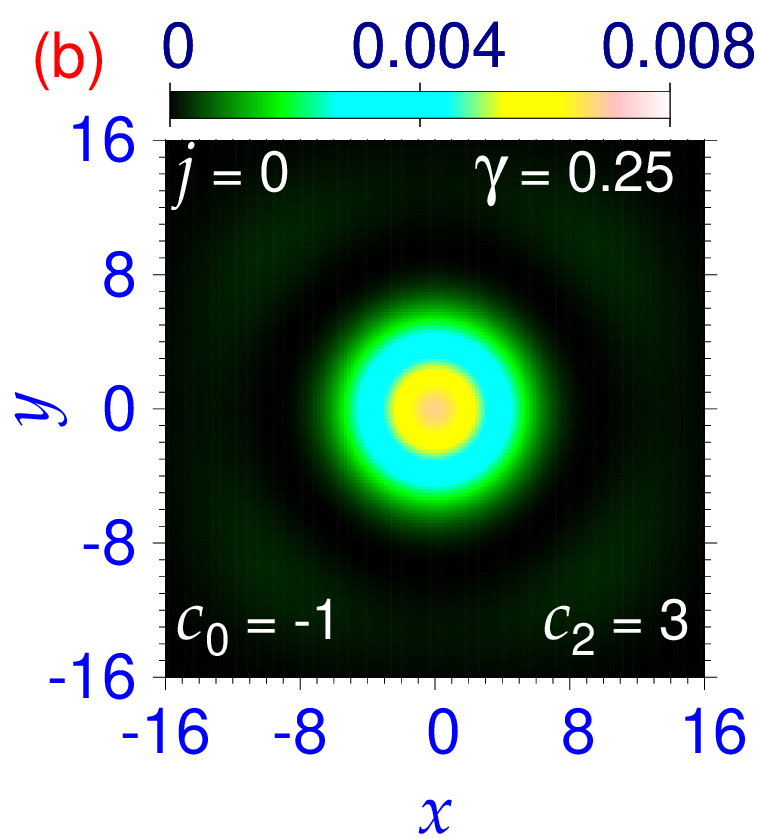}
\includegraphics[width=.325\linewidth]{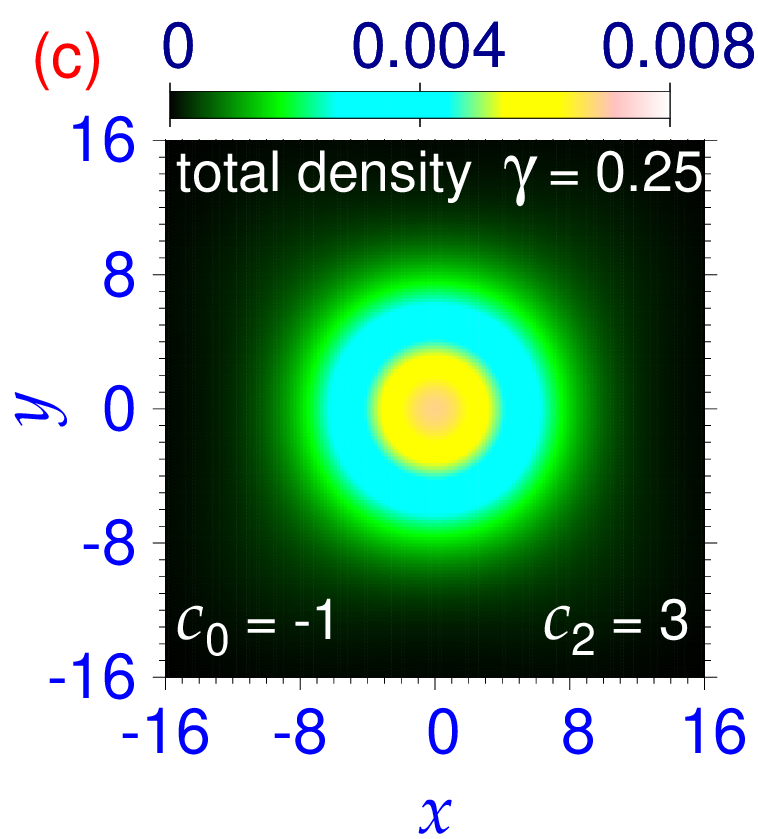}
 \includegraphics[width=.23\linewidth]{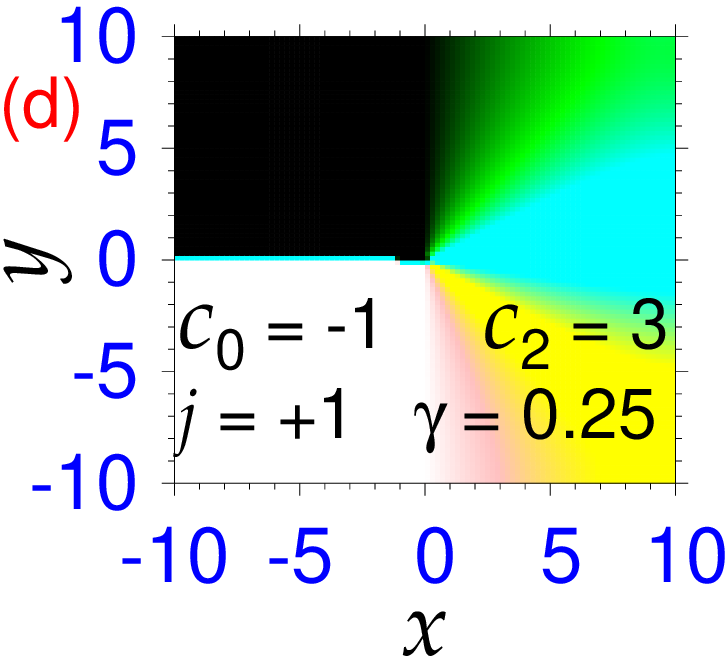} 
\includegraphics[width=.23\linewidth]{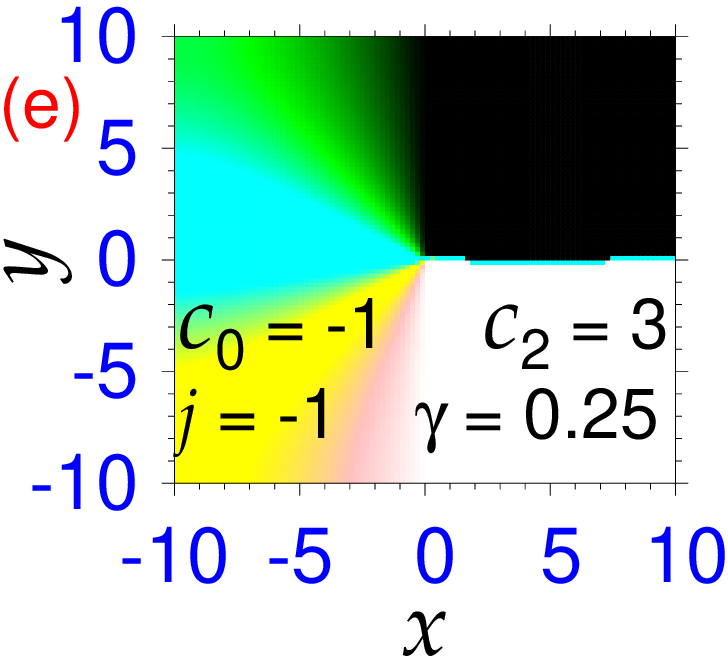}
 \includegraphics[width=.23\linewidth]{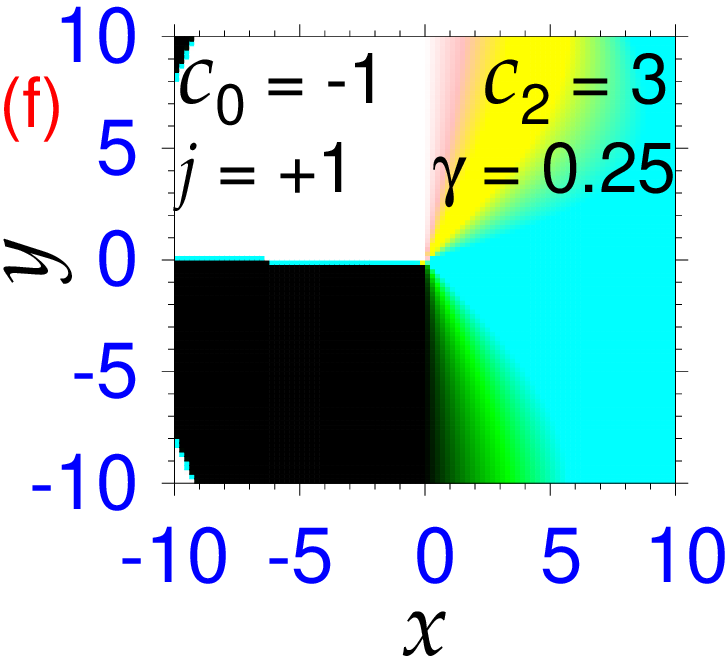}
 \includegraphics[width=.27\linewidth]{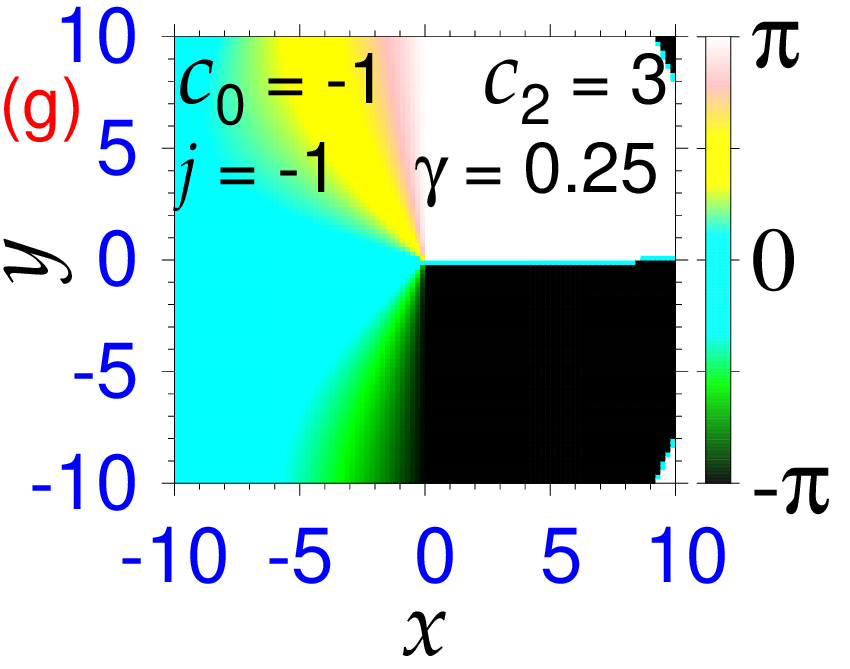} 

\caption{ Contour plot of density $n_j$ of a $(\mp 1,0,\pm 1)$-type spin-1 { Rashba or Dresselhaus SO-coupled} BEC soliton 
 for components (a) $j=\pm 1$, (b) $j=0$ and (c) total density. 
  The phases of the wave functions of components $j=\pm 1$ of the  $(-1,0,+1)$-type  Rashba SO-coupled  soliton 
shown in (a) are displayed in (d) and (e).  The same  of components $j=\pm 1$   for a $(+1,0,-1)$-type spin-1 Dresselhaus SO-coupled BEC soliton are illustrated in (f) and (g).  }
\label{fig2x}

\end{figure}

\begin{figure}[!t] 
\centering
\includegraphics[width=.325\linewidth]{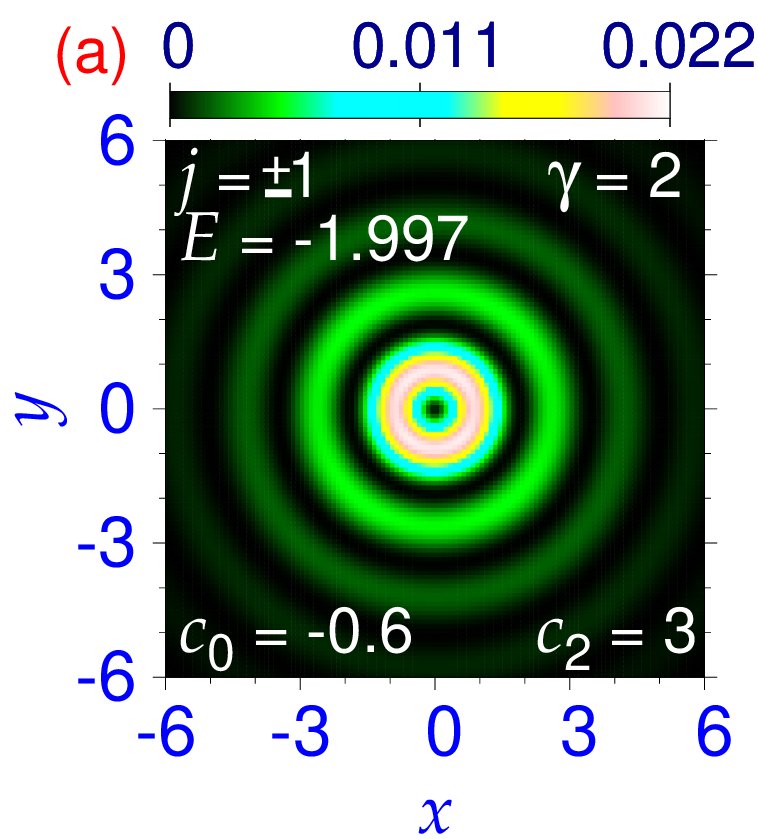} 
\includegraphics[width=.325\linewidth]{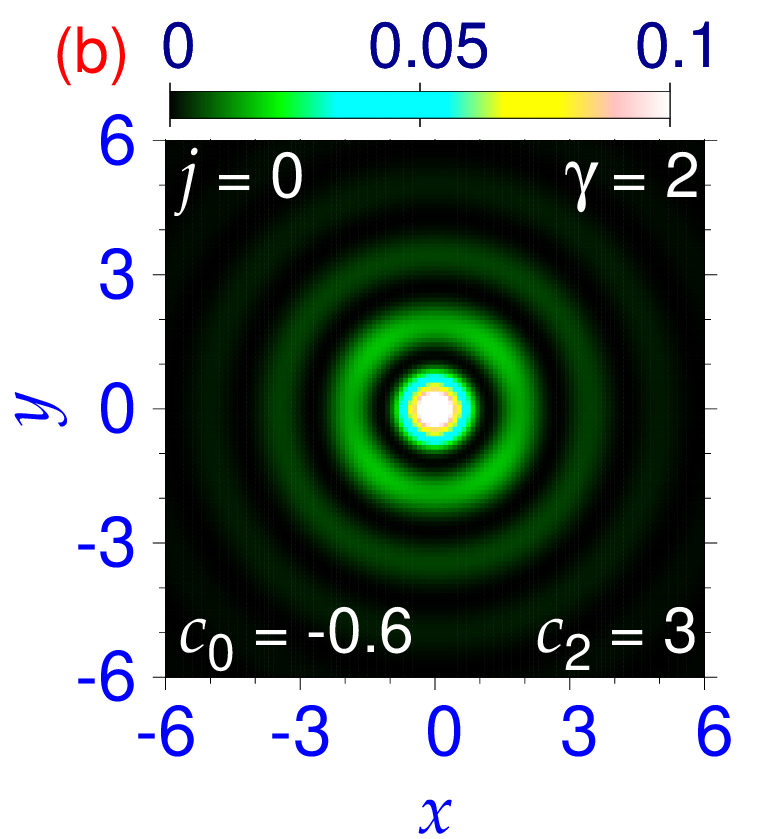}
\includegraphics[width=.325\linewidth]{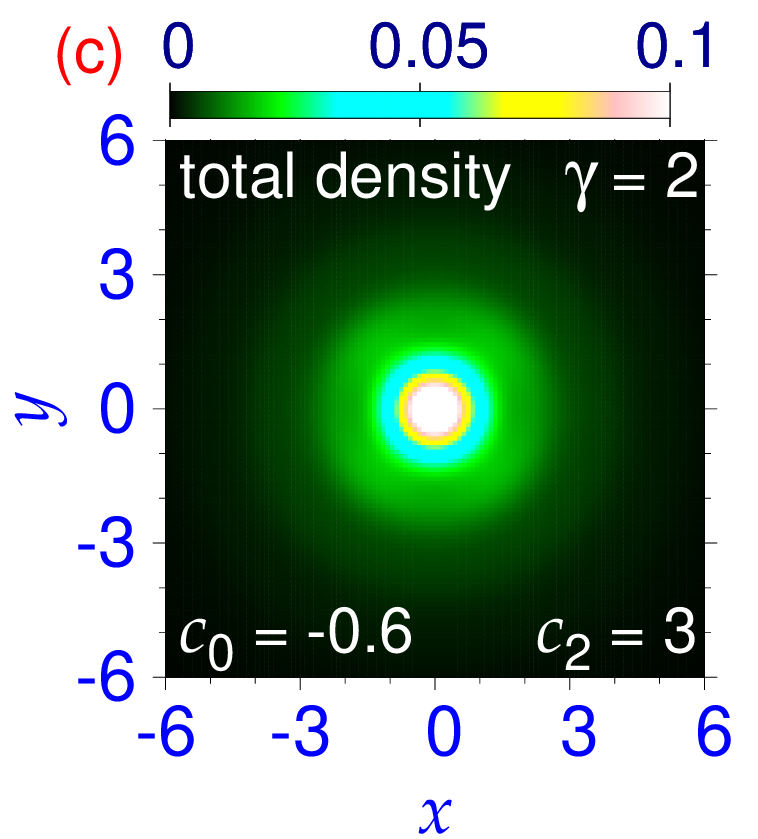}

\caption{ Contour plot of density  $n_j$ of a $(\mp 1,0,\pm 1)$-type multi-ring spin-1 { Rashba or Dresselhaus 
SO-coupled} BEC soliton 
 for components (a) $j=\pm 1$, (b) $j=0$  and (c) total density. }
\label{fig3}

\end{figure}

First, we consider $(\mp 1,0,\pm 1)$-type solitons for small $\gamma$ ($\gamma \lessapprox 0.75$) for Rashba  (upper sign) or Dresselhaus (lower sign) SO coupling.
In Fig. \ref{fig2x} we display the contour plot of   density  of components (a) $j=\pm 1$, (b) $j=0$, and (c) total density   of a  $(\mp 1,0,\pm 1)$-type soliton  for parameters $c_0=-1,c_2=3,\gamma=0.25$. The energy and density of the $(\mp 1,0,\pm 1)$-type soliton are insensitive to the variation of $c_2$ in the range $10>c_2>0$   \cite{quasi-1d12,quasi-2d1}.  
The energy of the $(\mp 1,0,\pm 1)$-type soliton in  Fig. \ref{fig2x}  is $E=-0.030$ independent of the type of SO coupling: Rashba or Dresselhaus.
  In Figs. \ref{fig2x}(d)-(e) we display the contour plot of the phase of wave function  components $j=\pm 1$ of the $(-1,0,+1)$-type Rashba SO-coupled 
soliton of Fig. \ref{fig2x}(a) showing a phase drop of $\mp 2\pi$ under a complete rotation, 
indicating angular momenta of $\mp 1$ in these components. Although the densities for Rashba and Dresselhaus 
SO-coupled solitons are the same the corresponding phases for Dresselhaus SO-coupling are different, viz. 
Figs. \ref{fig2x}(f)-(g) showing the phases of components $j=\pm 1$ of the Dresselhaus SO-coupled soliton
corresponding to a phase drop of  $\pm 2 \pi$  under a complete rotation,
indicating angular momenta of $\pm 1$  in these components. 
These phases are consistent with the vortex-antivortex structure 
of the $(\mp 1,0,\pm 1)$-type solitons for Rashba and Dresselhaus couplings.

To study   multi-ring solitons for medium  
 $\gamma$, 
in Fig. \ref{fig3} we show the contour plot of   density  of components (a) $j=\pm 1$ and (b) $j=0$  
and (c) total density 
of a  $(\mp 1,0,\pm 1)$-type multi-ring soliton  for Rashba or Dresselhaus SO couplings 
for $c_0=-0.6,c_2=3,\gamma=2$.    Although,  there is modulation in density of  different  components, the total density  shows no modulation. 
The phases in this case (not shown)  are identical to the same of the  $(\mp 1,0,\pm 1)$-type
soliton shown in Figs. \ref{fig2x}(d)-(g) reflecting the same vortex-antivortex structure  
at the center of the respective components for Rashba or Dresselhaus SO couplings. 
%{\color{green} The  $(-1,0,+1)$-type multi-ring solitons of Figs. \ref{fig3} appear for both positive and negative values  $c_2$ $(c_2>c_2^{(\mathrm{cr})} <0)$, thus covering both 
%ferromanetic and antiferromagnetic phases.}   
 The energy of the $(\mp 1,0,\pm 1)$-type multi-ring soliton in  Fig. \ref{fig3} is    $E=-1.997$ for both SO couplings. 
  The increase of $\gamma$ from Fig.  \ref{fig2x}  to  Fig.  \ref{fig3} has increased the binding, 
and hence aids in  forming soliton. 
The change in binding due to the change of $c_0$ from Fig.  \ref{fig2x}  to  Fig.  \ref{fig3} is negligible 
in this scale.
  Multi-ring solitons were also investigated in a quasi-2D pseudo spin-1/2 SO-coupled BEC trapped in a radially periodic potential \cite{radper} which creates the multi-ring modulation in density. However,  the present radial modulation in density without any external trap 
is a consequence of the SO couplings.

\begin{figure}[!t] 
\centering

\includegraphics[width=.325\linewidth]{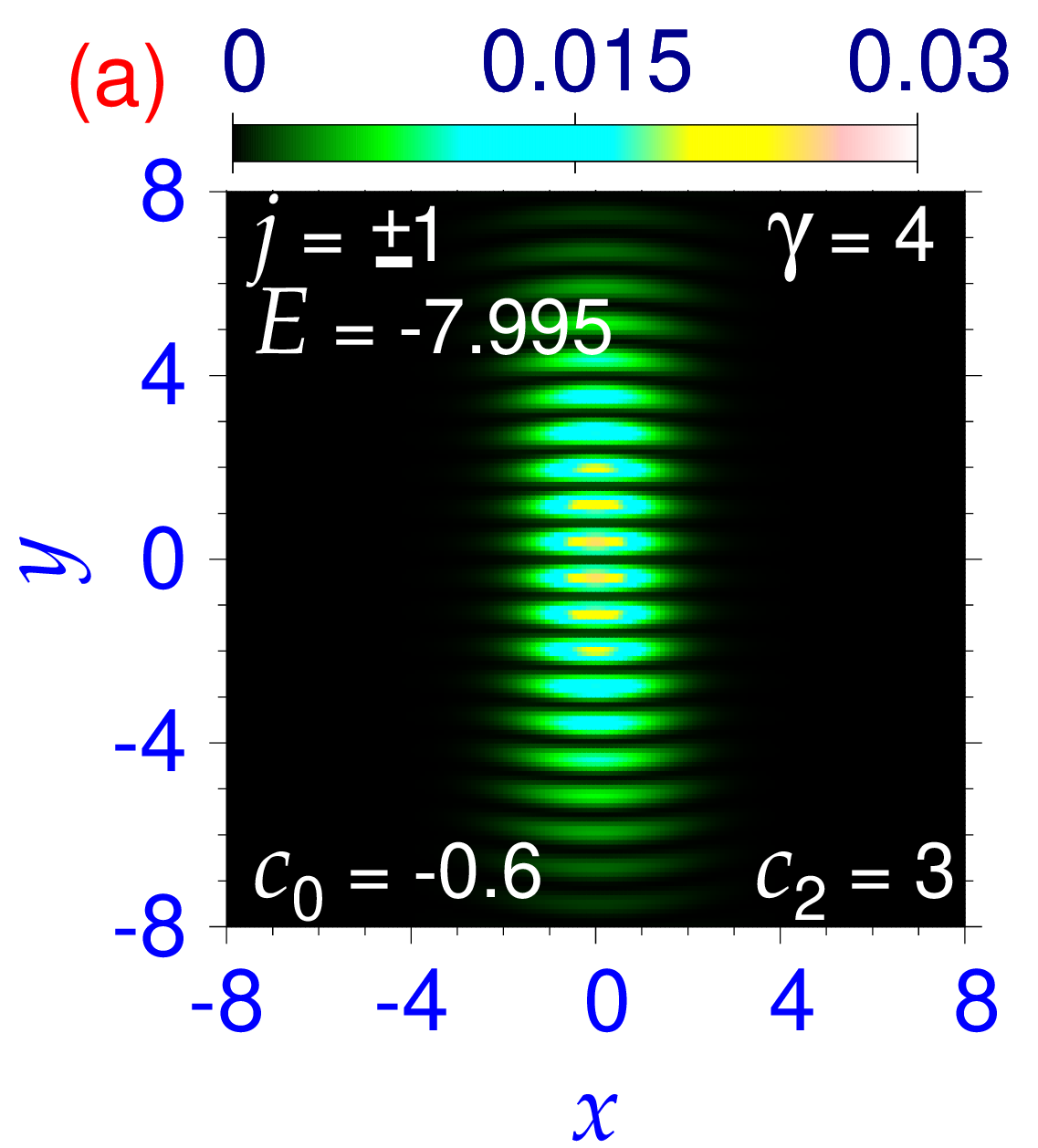}
\includegraphics[width=.325\linewidth]{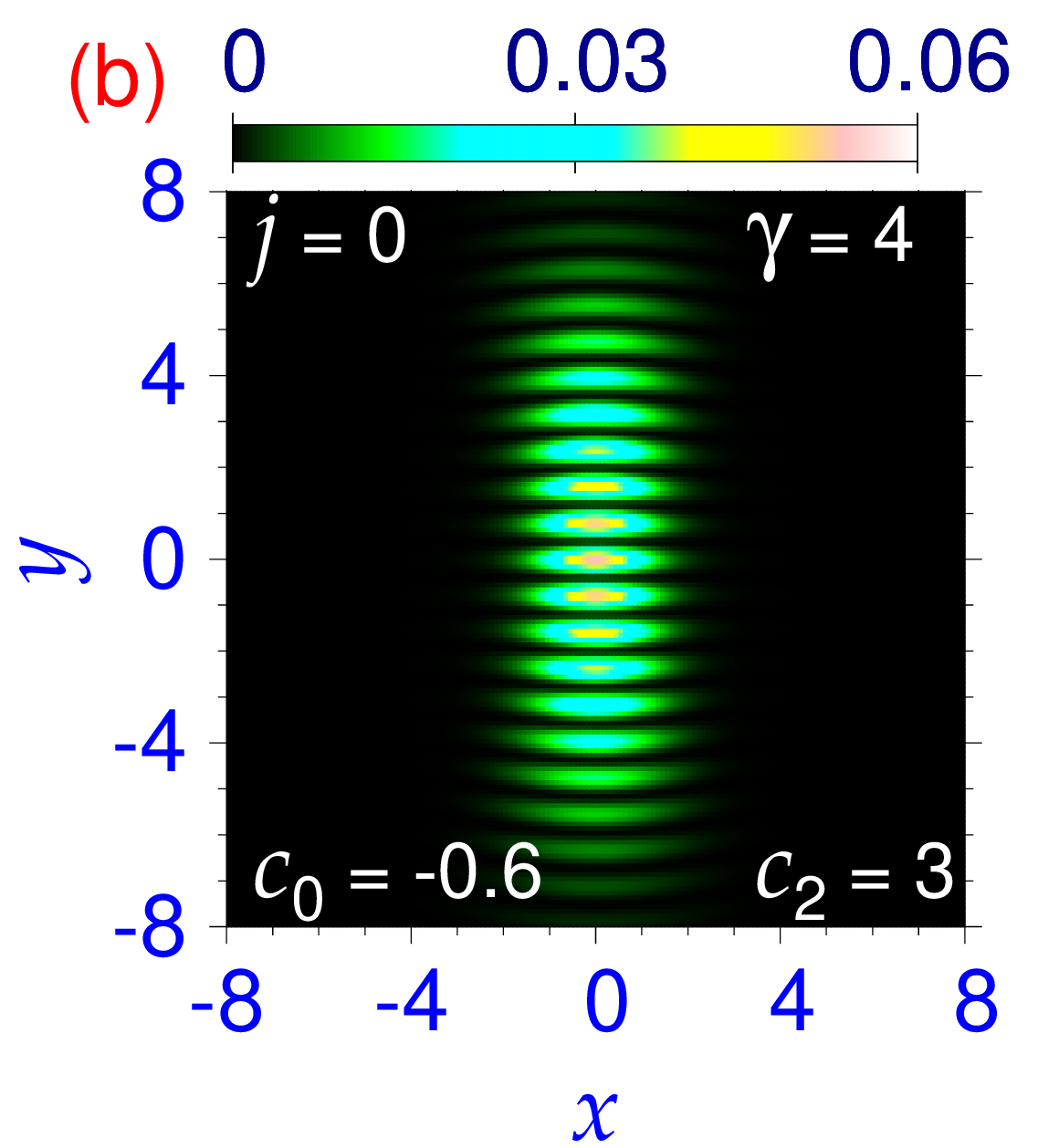}
 \includegraphics[width=.325\linewidth]{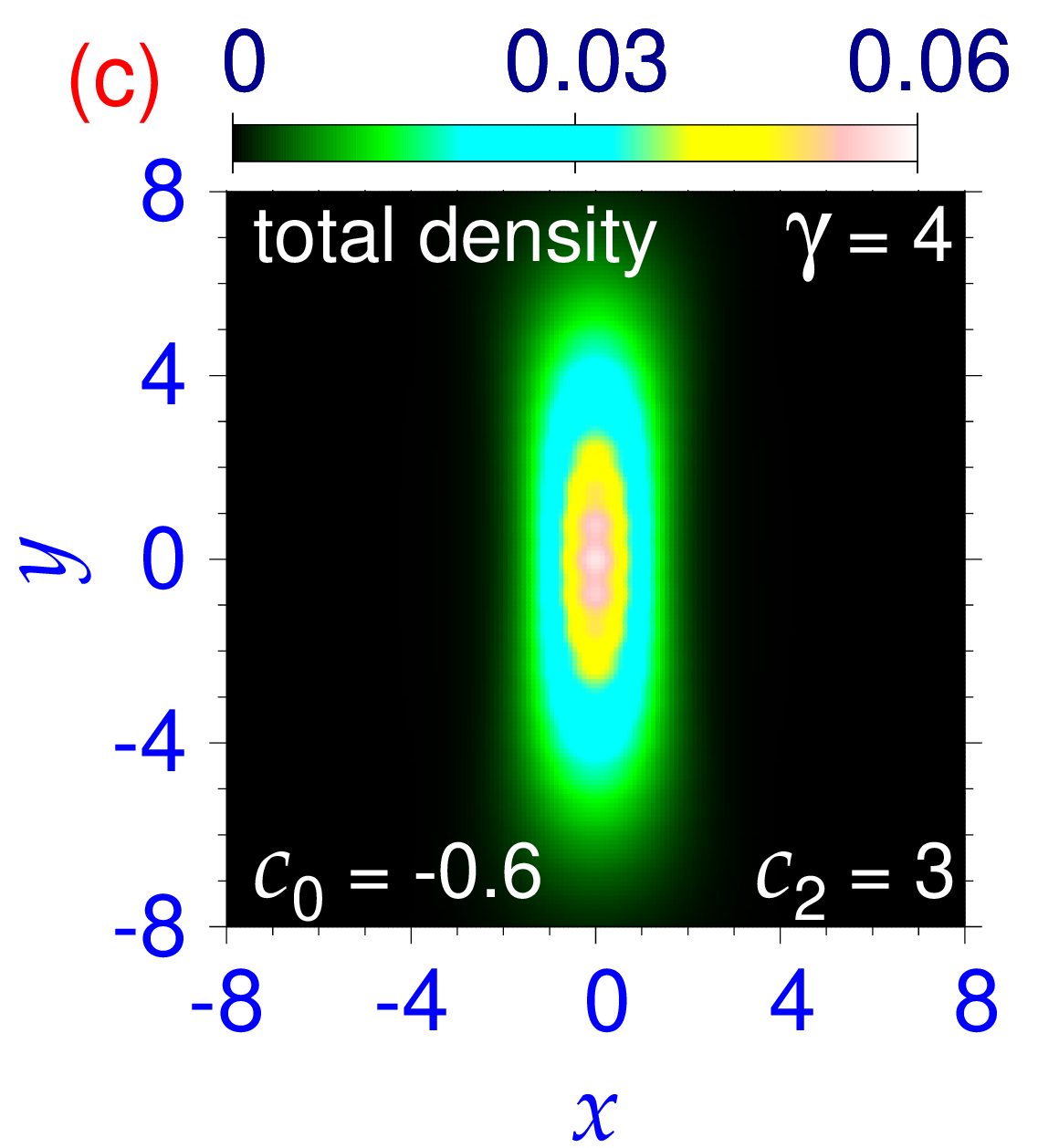} 

\includegraphics[width=.325\linewidth]{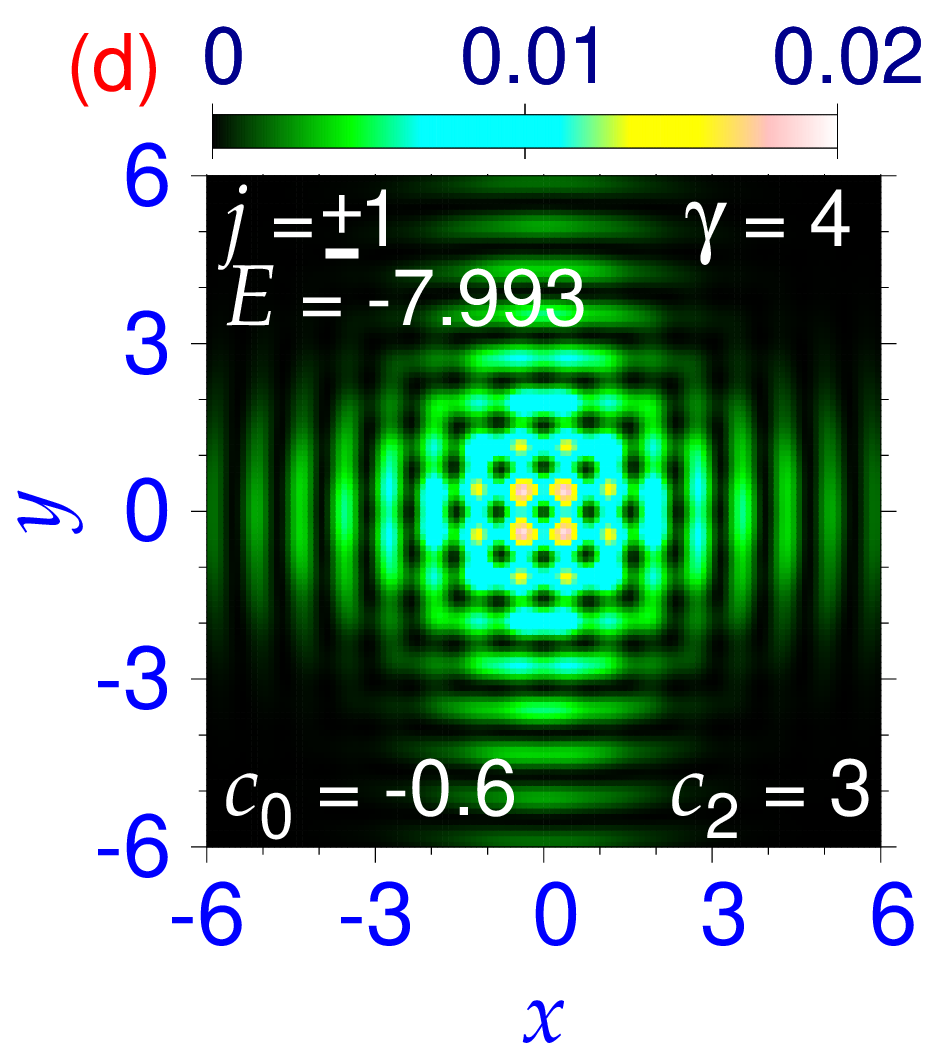}
\includegraphics[width=.325\linewidth]{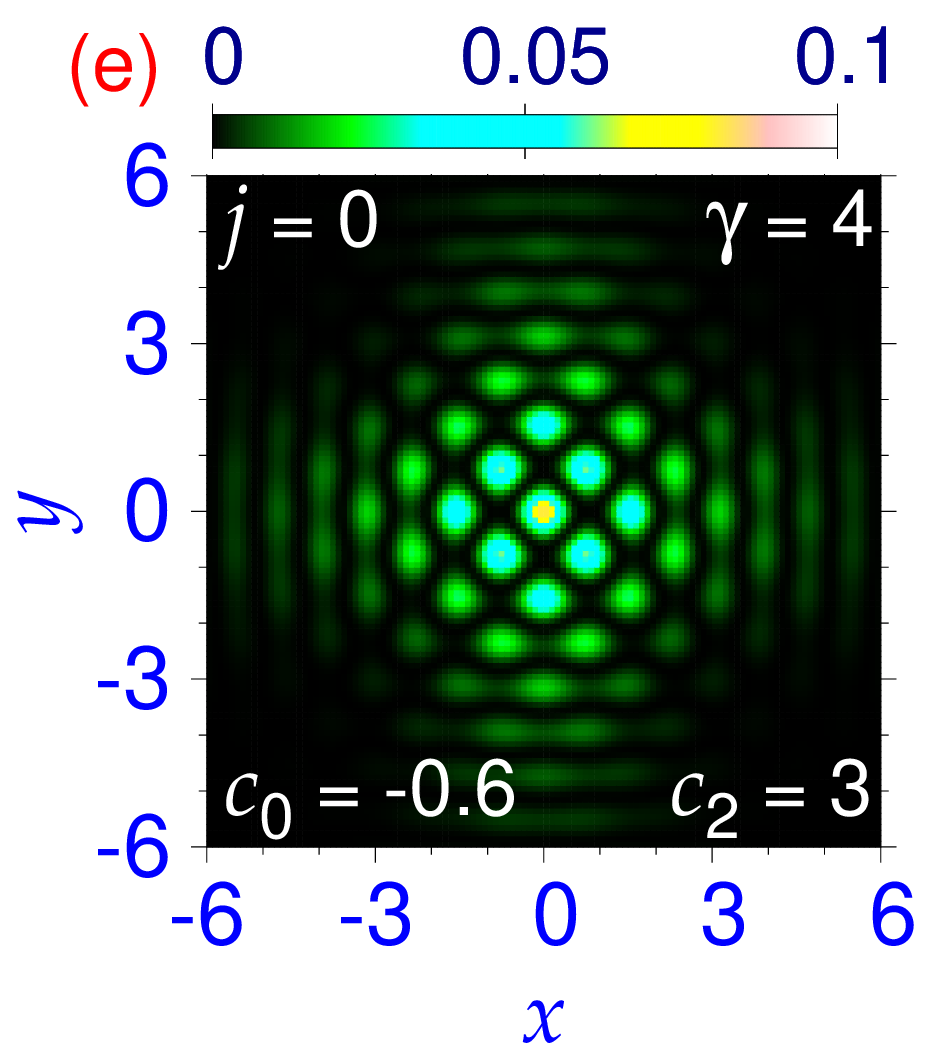} 
 \includegraphics[width=.325\linewidth]{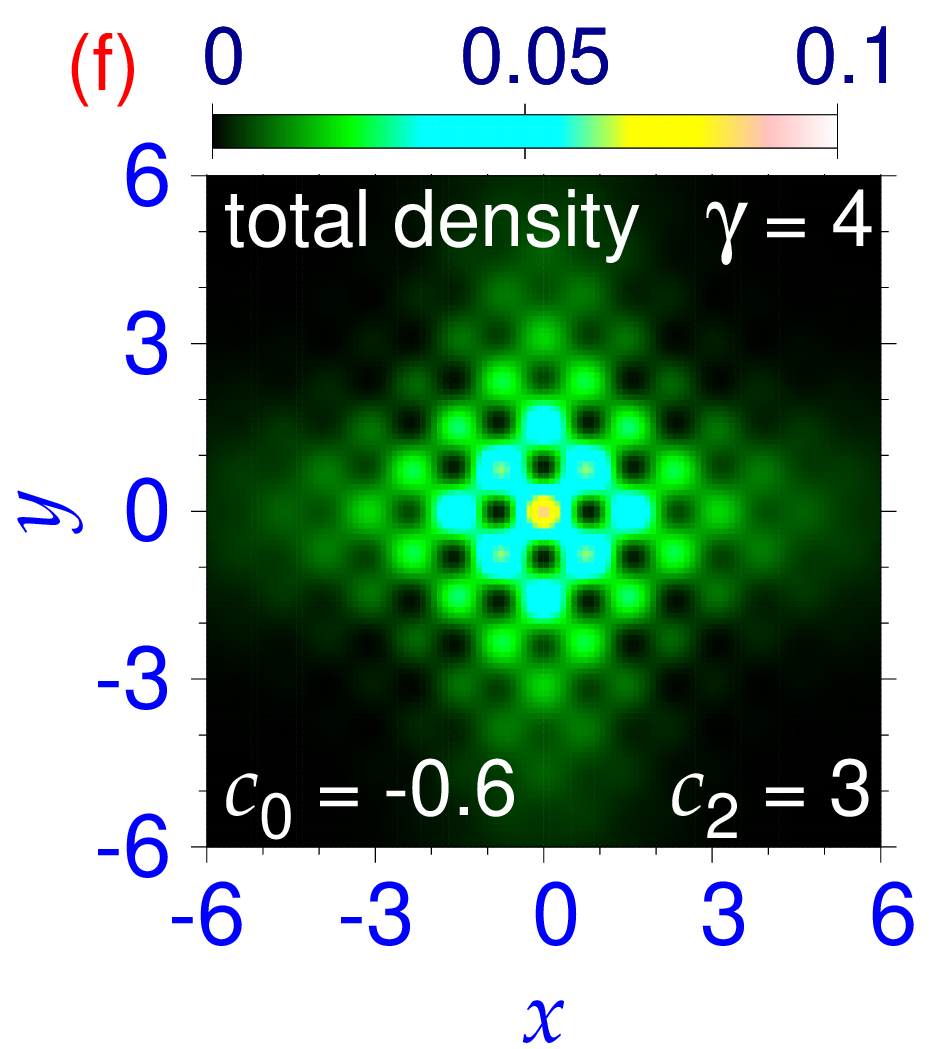} 

\caption{   Contour plot of density $n_j$ of a stripe spin-1 { Rashba  or Dresselhaus SO-coupled} BEC soliton 
 of components (a) $j=\pm 1$ and (b) $j=0$ { (c) the total density}.  
The same of a  super-lattice  soliton 
  of components {(d) $j= \pm 1$,  (e) $j=0$, and (f) the total density}. }
\label{fig4}

\end{figure}

{As $\gamma$ is increased,  the    $(\mp 1,0,\pm 1)$-type
multi-ring solitons 
continue to exist but will not be considered further. Two new
types of solitons {\it without} any vortex at the center of the components appear as quasidegenerate states: stripe and super-lattice solitons with periodic distribution of matter in $x$ and/or $y$ directions.
The single particle Hamiltonian (\ref{sp}) should have solutions of  the plane wave form $\exp(\pm i\alpha \gamma x)\otimes \exp (\pm i\beta \gamma y)$, where $\alpha$ and $\beta$ are constants. In  the presence of interaction ($c_0,c_2 \ne 0$), the solution will be a superposition of such plane wave solutions leading to a periodic variation of density in the form $\sin^2(\alpha \gamma x)$, $\cos^2(\beta \gamma y)$, $\sin^2(\alpha \gamma x)\sin^2(\beta \gamma y)$ etc. appropriate for stripe or lattice solitons. This has been demonstrated in details for quasi-1D solitons \cite{quasi-1d1}, the same of  the present quasi-2D solitons will be the subject of a future investigation. The period of the lattice or stripe increses as $\gamma$ is reduced. For small $\gamma$, the size of the soliton is smaller than this period and the periodic pattern in density  is not possible.}
 In Fig. \ref{fig4} we illustrate a quasi-2D  stripe soliton for  $c_0=-0.6, c_2 =3,$ and $\gamma=4$ through a contour plot of density of components (a) $j=\pm 1$, (b) $j=0$, and (c) total density, obtained by imaginary-time propagation using an initial localized wave function  modulated by appropriate stripes in form $\sin(\gamma x)$ and $\cos(\gamma x)$  for $j=\pm 1$  and 0 components. 
Although, there is a stripe pattern in density in this case,
the positions of maxima in components $j=\pm 1$ coincide with the minima in component $j=0$ due to a phase separation among the components, thus resulting in a total density without modulation.
 In Fig. \ref{fig4}
the super-lattice soliton for the same set of parameters 
is displayed through a contour plot of density of components (d) $j=\pm  1$, (e) $j=0$, and (f)  total density,  obtained by imaginary-time propagation using the converged wave function of Fig. \ref{fig3} as the initial state. 
%In  Figs.   \ref{fig4}(f)-(h) the super-lattice soliton for the  parameters $c_0=-0.4, c_2 =3, m=0$ and $\gamma=4$ are exhibited.   
The distribution of matter on a 2D square lattice is prominent in the total density plot of 
Fig.  \ref{fig4}(f)  \cite{2020}.
%The present super-lattice soliton  is a consequence of the SO coupling and breaks {\it continuous}  translational symmetry as required in a super-solid \cite{sprsld}.  
%The super-lattice soliton in Figs. \ref{fig4}(f)-(h) with $c_0=-0.4$ has reduced attraction compared to the one
%in Figs. \ref{fig4}(c)-(e) with $c_0=-0.6$, consequently,  has larger size and more prominent and clear lattice structure, viz.  Figs. \ref{fig4}(g)-(h). 
For the same  set of parameters,  the stripe soliton of energy  $E=-7.995$
and the super-lattice soliton of energy  $E=-7.993$
are almost degenerate states 
with the stripe solitons having slightly smaller energy.

 \begin{figure}[!t]
\centering 
\includegraphics[width=.325\linewidth]{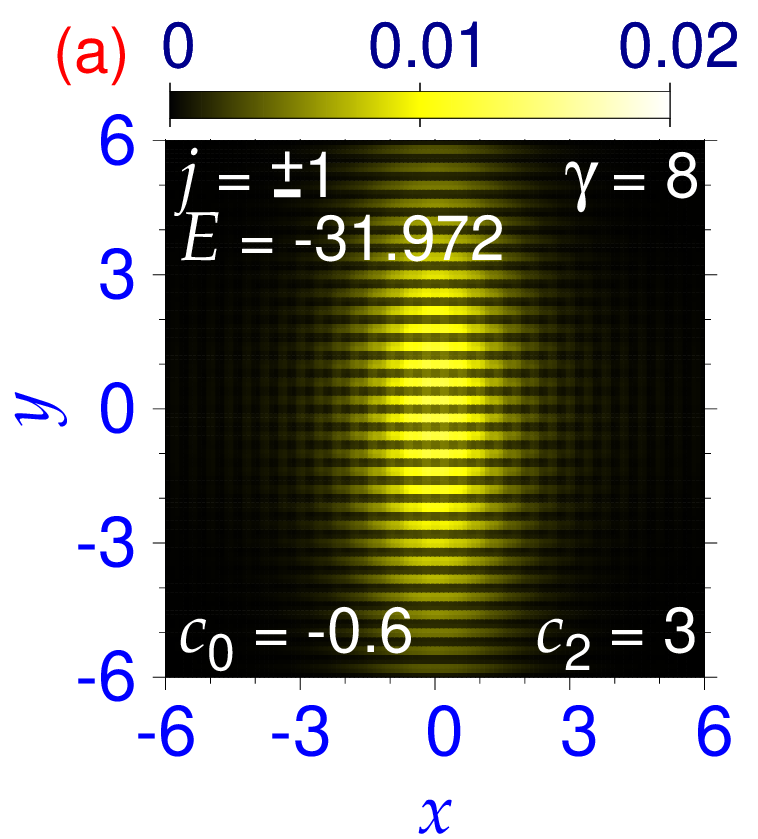}
 \includegraphics[width=.325\linewidth]{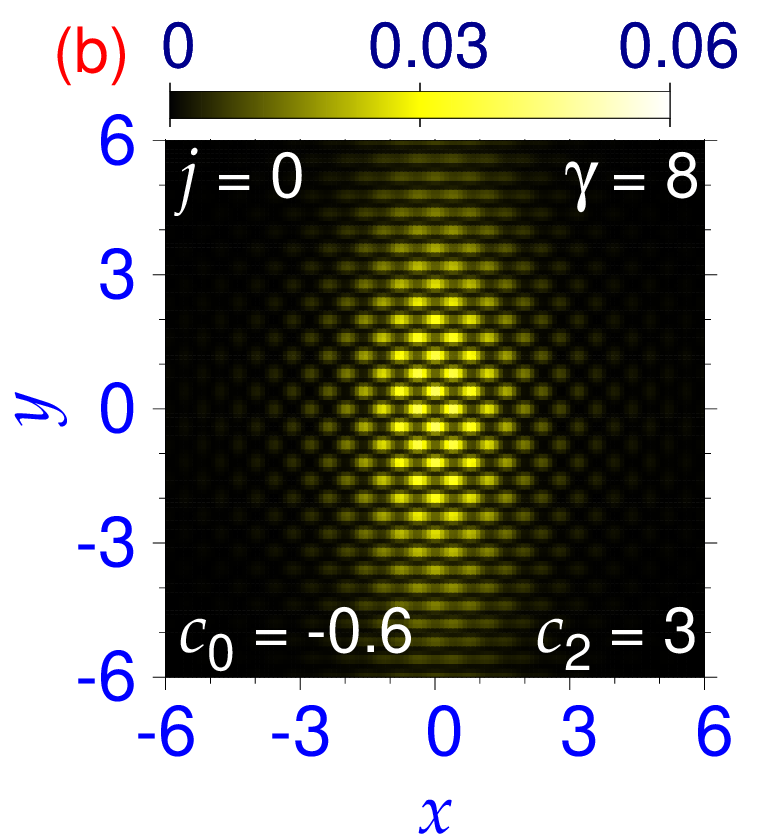} 
\includegraphics[width=.325\linewidth]{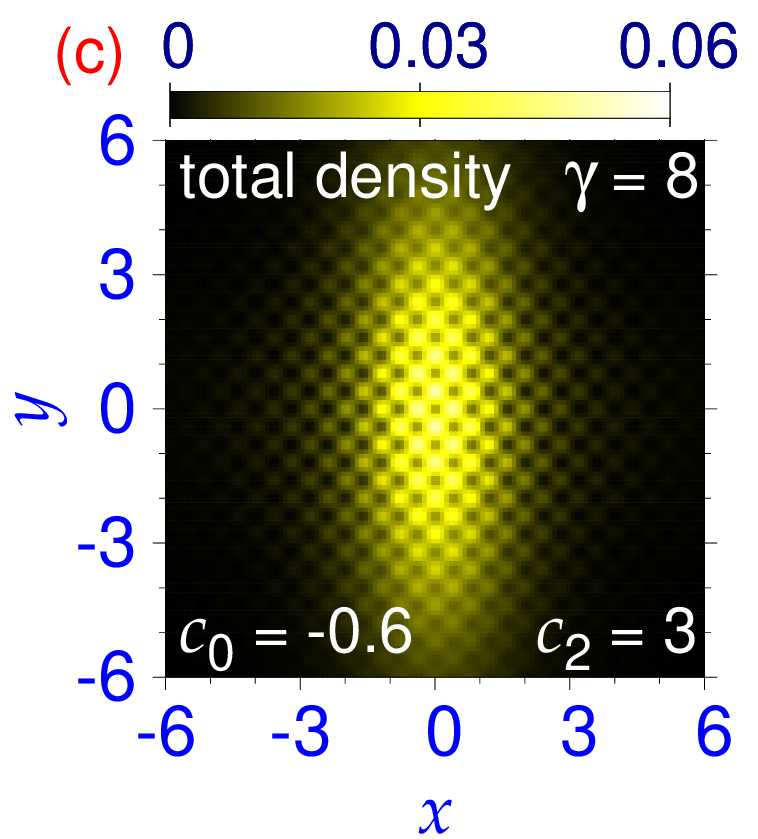}
 \includegraphics[width=.325\linewidth]{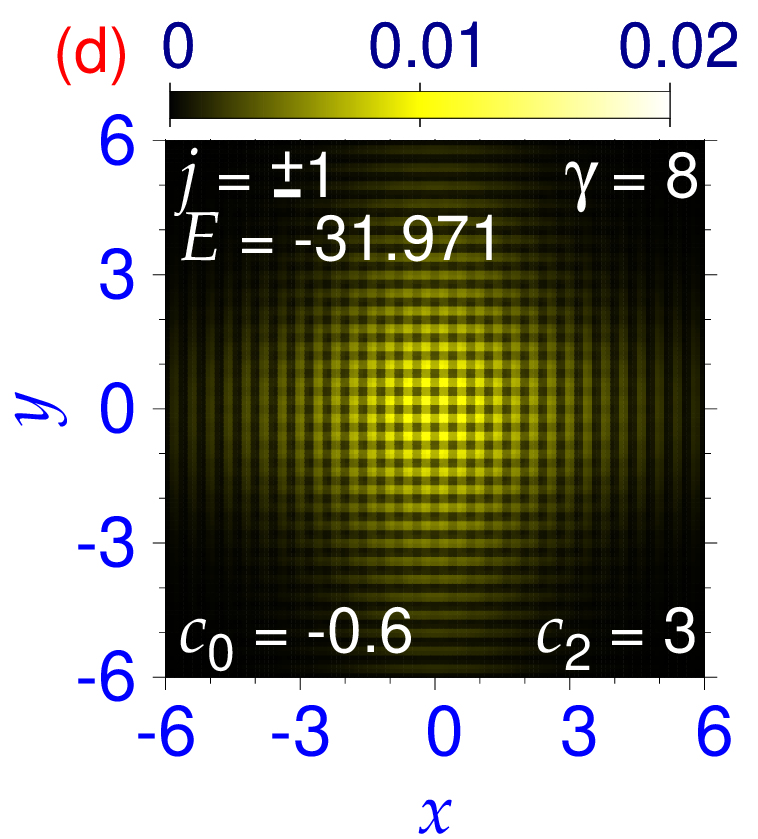} 
\includegraphics[width=.325\linewidth]{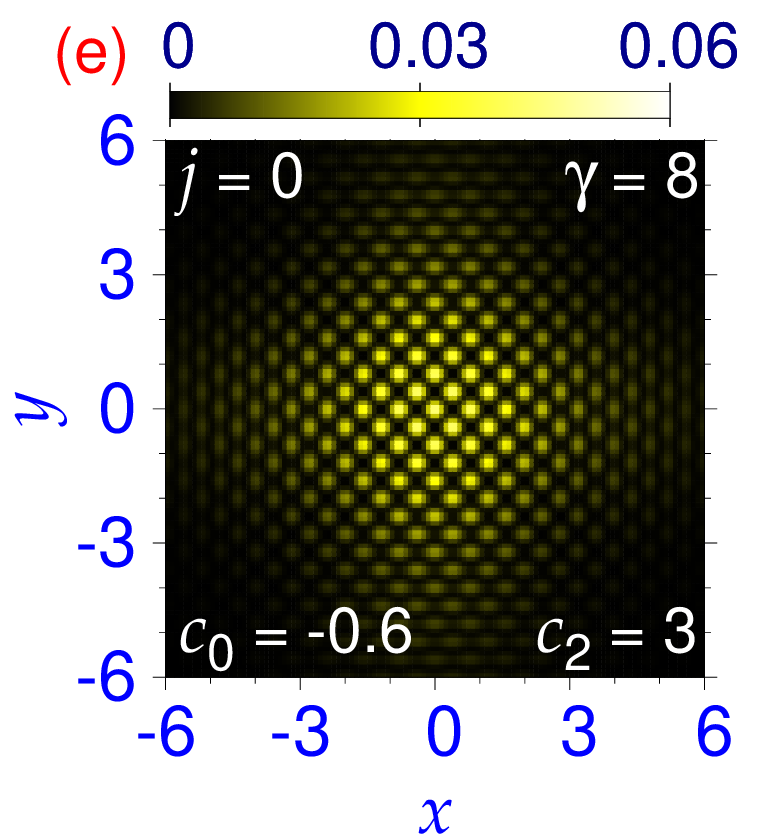} 
 \includegraphics[width=.325\linewidth]{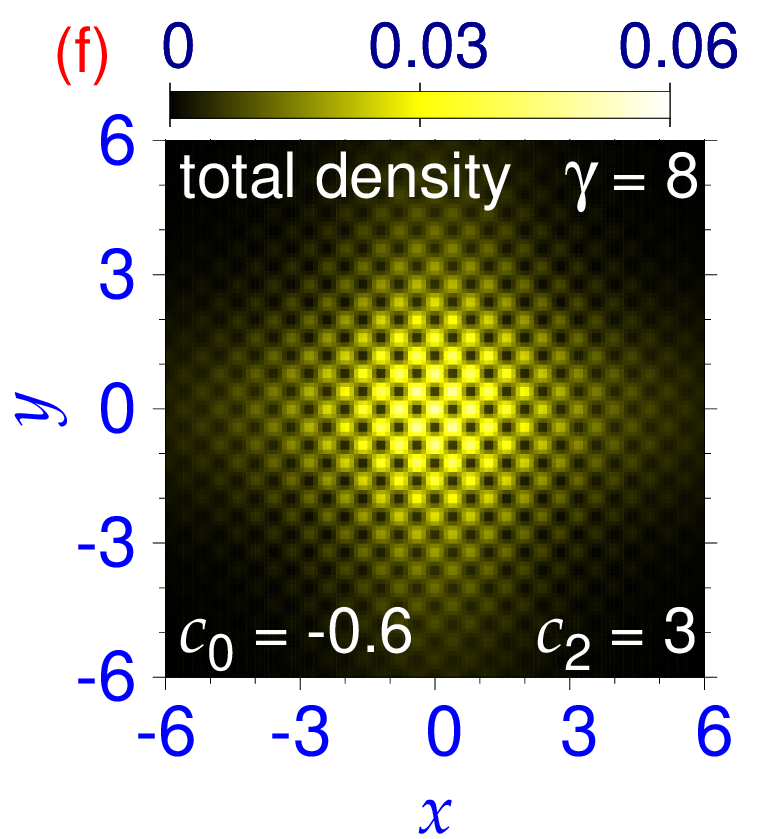}

\caption{ Contour plot of density $n_j$  of a super-lattice spin-1 { Rashba or Dresselhaus SO-coupled} BEC soliton (with a periodic stripe  in  
$j=\pm 1$)
 for components {(a) $j=\pm  1$, (b) $j=0$, and (c) the total density}.  
The same of a  super-lattice  soliton with a square-lattice pattern
 in components {(d) $j= \pm 1$,  (e) $j=0$,    and (f) the total density}.  }
\label{fig5}
\end{figure}

As $\gamma$ is further increased, the stripe solitons become super-lattice solitons with 2D square-lattice structure in total density. 
  In Fig. \ref{fig5} we display the contour plot of density  
of such a super-lattice soliton of components (a) $j=\pm 1$, (b)  $j=0$, (c) total density   for parameters  $c_0=-0.6, c_2=3$,  and $\gamma=8$.
{In this case, the component densities of Fig. \ref{fig5}(a)  $(j=\pm 1)$
continue to possess a 1D stripe pattern  but   
the total density in Fig. \ref{fig5}(c)  and { the density of the $j=0$ component in Fig. \ref{fig5}(b) 
develop  a prominent 2D square-lattice structure \cite{2020}.} 
The component densities as well as the 
total density  of a super-lattice  soliton with 2D square-lattice structure in both  
are displayed in 
 Fig. \ref{fig5}  for (d) $j=\pm 1$, (e)  $j=0$,  and (f) the total density. }
These two types of  super-lattice solitons have practically the same 
energy:  $E= -31.972$ and $-31.971$, respectively.

To demonstrate the dynamical stability of the solitons, we consider 
the  stripe and the super-lattice solitons of Fig. \ref{fig4} and subject the corresponding imaginary-time wave functions  to real-time propagation during 100 units of time after changing $c_2$ from 3 to 4.  The resultant density of the 
stripe  soliton is displayed in Fig. \ref{fig6x}
for 
 components (a) 
$j=\pm 1$, (b)  $j=0$, and  (c) total density.  
The same for the super-lattice
soliton    are shown in  Figs. \ref{fig6x}(d)-(f). 
%These densities are in good agreement with the corresponding densities of Fig.  \ref{fig4}, demonstrating the dynamical stability. 
Although the root-mean-square sizes and energy were oscillating (different energies in Figs. \ref{fig4} and \ref{fig6x}) during real-time propagation, the periodic pattern in total density survived at $t=100$. If the solitons were not
dynamically  stable, the small numerical deviation due to a change of   $c_2$ from 3 to 4 would have destroyed the periodic pattern in density.

 \begin{figure}[!t]
\centering 
\includegraphics[width=.325\linewidth]{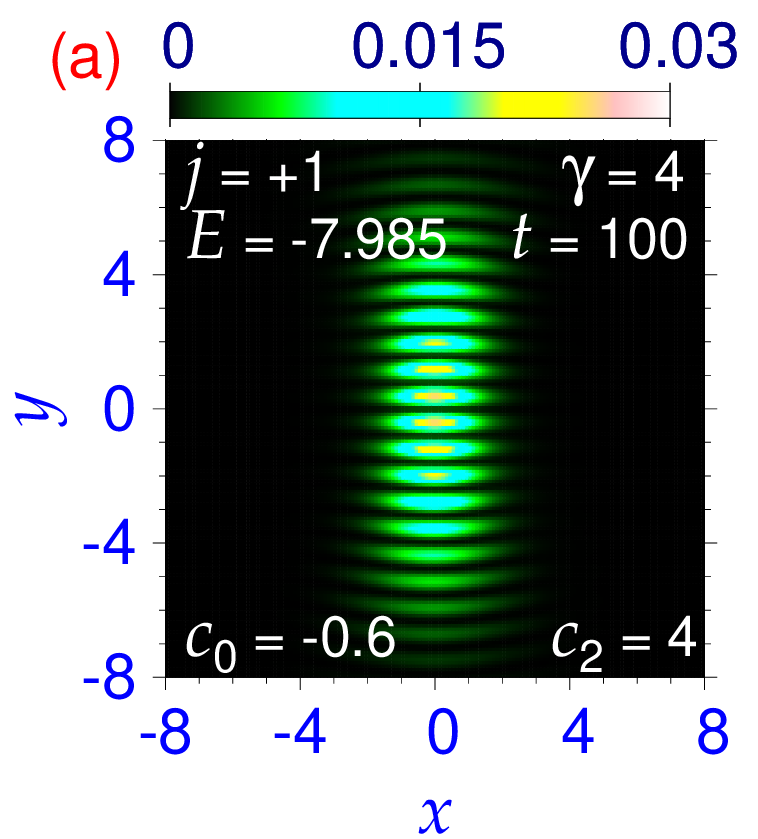}
 \includegraphics[width=.325\linewidth]{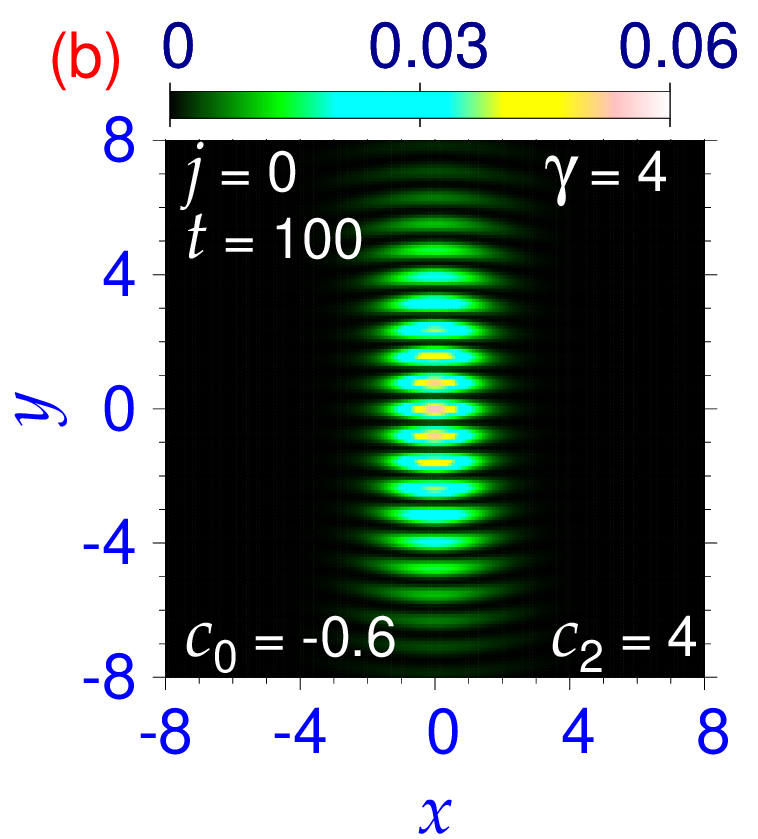} 
\includegraphics[width=.325\linewidth]{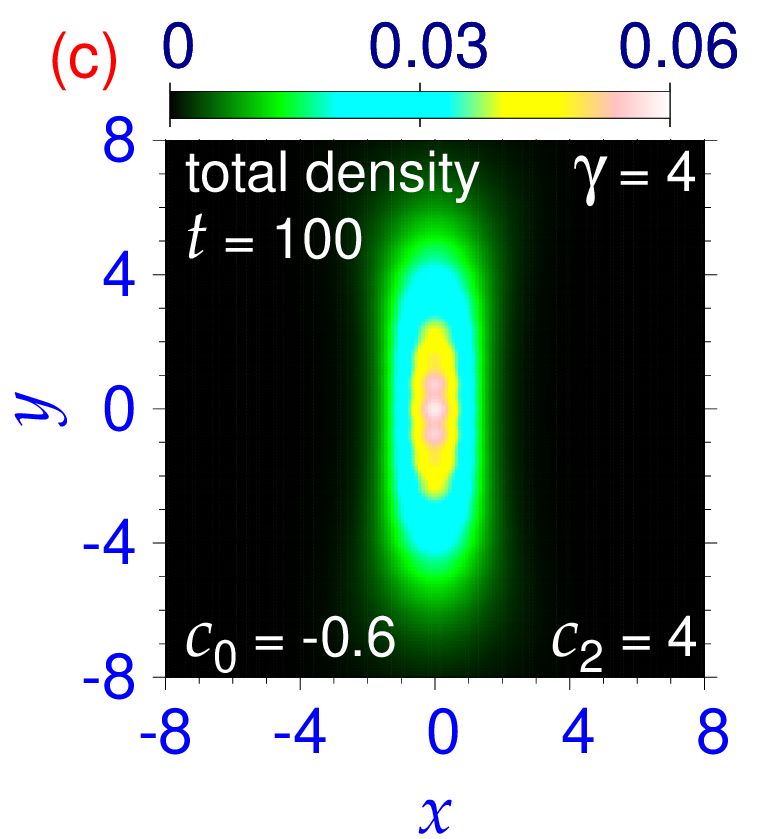}
 \includegraphics[width=.325\linewidth]{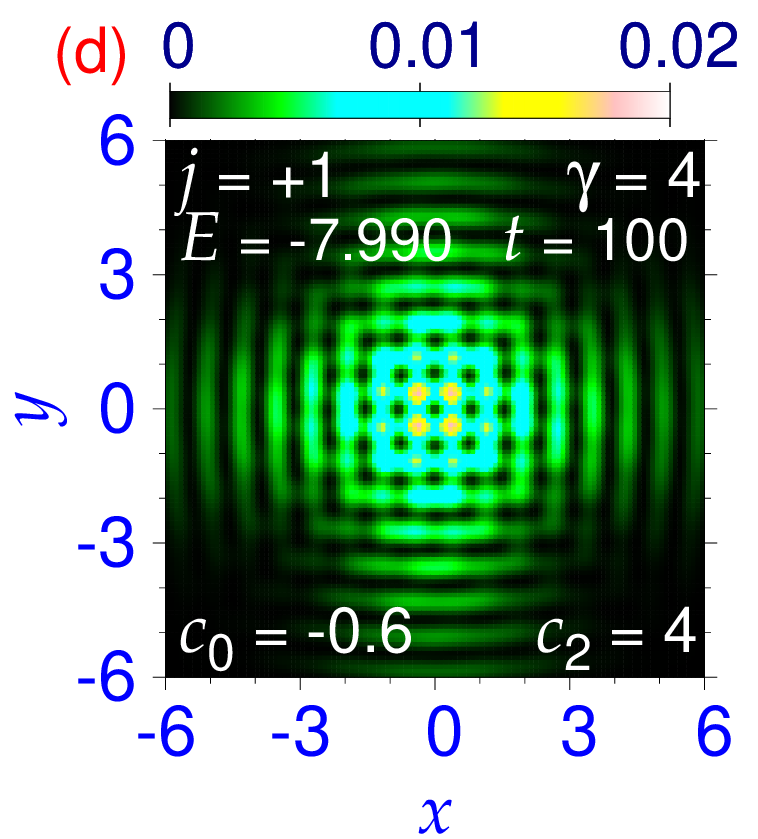}  
\includegraphics[width=.325\linewidth]{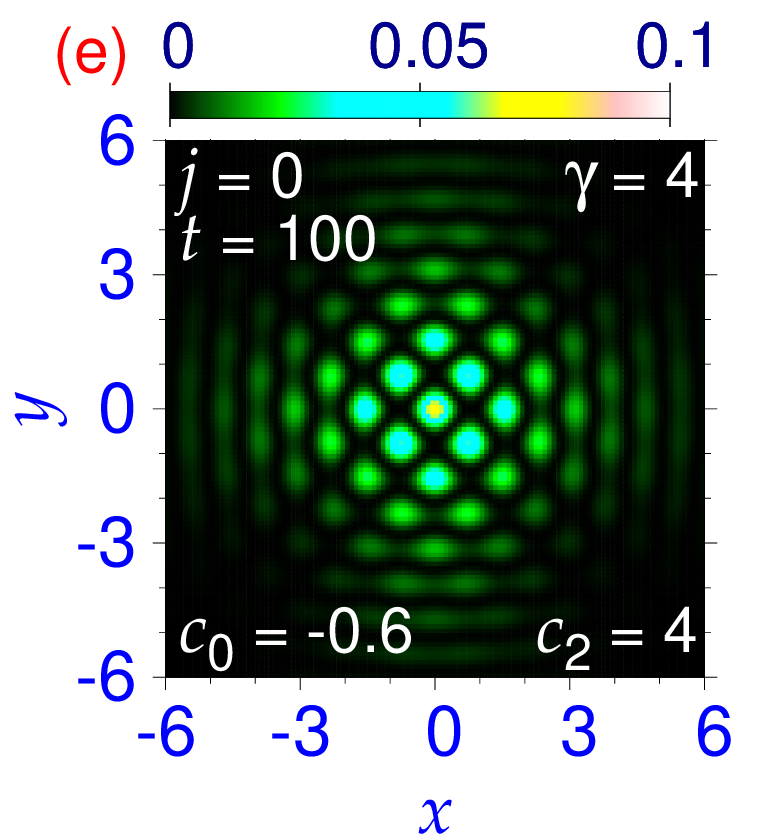}
 \includegraphics[width=.325\linewidth]{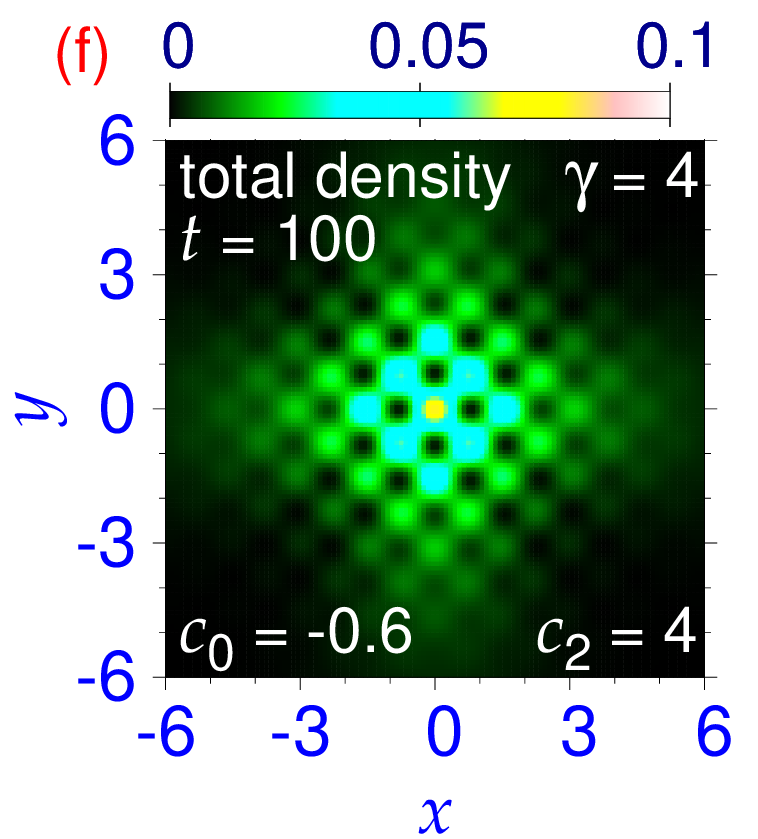}

\caption{ Contour plot of density of a stripe  BEC soliton,  of Fig.  \ref{fig4}(a)-(c), 
 of components {(a) $j=\pm 1 $, (b) $j=0$ and (c) the total density}, after real-time propagation at time $t=100$  upon a change of $c_2$ from 3 to 4 at $t=0$.  
The same of a  super-lattice  soliton, of  Fig.  \ref{fig4}(d)-(f), of components
{
 (d) $j= \pm 1$,  (e) $j=0$ and (f) the total density.}  }
\label{fig6x}
\end{figure}

{ At a finite temperature, there will be a thermal cloud outside the super-fluid 
soliton and the present study at zero temperature may  require corrections \cite{rmp}.
For  ultra-low-density scalar BEC solitons  \cite{r2,r3}, the experimental result \cite{agree} for the critical number of atoms in a   $^{85}$Rb soliton 
at low temperature
is in excellent  agreement with that 
derived from the zero-temperature GP equation \cite{arnaldo} with an estimated error of less than $1\%$.
As the present solitons are also of ultra-low density with very small $|c_0|$, the finite-temperature  corrections in an experiment
are expected to be small. Hence we do not believe the finite-temperature correction to be so large as to invalidate the general findings of this study. }

{In the pseudo spin-1/2 SO-coupled quasi-2D BEC only   stripe  state was found \cite{sinha}. 
In the quasi-2D spin-1 case, in this study, we find both stripe and 2D square lattice states.
However, in a quasi-2D spin-1 BEC we could not find any super-lattice   state for an equal mixture of 
Rashba and Dresselhaus couplings, which is a simpler 1D coupling. 
Preliminary studies indicate that in the quasi-2D spin-2 case there could be formation of stripe, square
lattice and hexagonal lattice \cite{sandeep} states. It seems that a  more intricate SO coupling with  an increased number of 
wave function components is responsible for a diverse type of  super-lattice states.
}

We demonstrated new types of dynamically stable solitons in { a Rashba or Dresselhaus} SO-coupled spin-1 quasi-2D untrapped   spinor {antiferromagnetic} ($c_2>0$) BEC  
for medium and large
SO-coupling strengths $\gamma$.    
 For medium $\gamma$,  $(\mp 1,0,\pm1)$-type 
multi-ring solitons are found for Rashba and Dresselhaus SO couplings.  
As $\gamma$ is increased,  stripe and  super-lattice 
solitons  are found, where matter is  distributed  either  in stripe form with maxima and minima
or on a square quasi-2D lattice. The total density of the stripe soliton has no density modulation, 
whereas that of the super-lattice soliton shows a 2D square lattice pattern,  viz. Fig. \ref{fig4} for $\gamma=4$. As $\gamma$ is further increased, the stripe solitons become super-lattice solitons,  
 viz. Figs. \ref{fig5}(a)-(c) for $\gamma=8$, where the total density exhibits a 
2D periodic structure on a square lattice.
{ The spontaneous  formation of a 2D square-lattice structure in the total density of
super-lattice solitons is remarkable in being a trap-less super-fluid system  
   showing   crystallization behavior.}   These super-lattice solitons are dynamically robust and deserve   further  theoretical and experimental  investigation.

\begin{acknowledgments}
The author thanks Dr. Sandeep Gautam and Ms. Psrdeep Kaur for discussion and 
 acknowledges support by the CNPq (Brazil) grant 301324/2019-0, and by the ICTP-SAIFR-FAPESP (Brazil) grant 2016/01343-7

\end{acknowledgments}

\end{document}